\documentclass[pre,twocolumn,aps,eqsecnum]{revtex4}
\usepackage{amsmath,bm,epsfig}
\newcommand{\B}[1]{{\bm{#1}}}

\begin{document}

\title{Nonequilibrium Thermodynamics of Amorphous Materials II:\\
 Effective-Temperature Theory}
\author{Eran Bouchbinder}
\affiliation{Racah Institute of Physics, Hebrew University of Jerusalem, Jerusalem 91904, Israel}
\author{J. S. Langer}
\affiliation{Dept. of Physics, University of California, Santa Barbara, CA  93106-9530}


\begin{abstract}
We develop a theory of the effective disorder temperature in glass-forming materials driven away from thermodynamic equilibrium by external forces. Our basic premise is that the slow configurational degrees of freedom of such materials are weakly coupled to the fast kinetic/vibrational degrees of freedom, and therefore that these two subsystems can be described by different temperatures during deformation. We use results from the preceding paper on the nonequilibrium thermodynamics of systems with internal degrees of freedom to derive an equation of motion for the effective temperature and to learn how this temperature couples to the dynamics of the system as a whole.
\end{abstract}

\maketitle

\section{Introduction}
\label{intro}

The effective disorder temperature is emerging as an essential ingredient in theories of nonequilibrium phenomena in amorphous materials \cite{CUGLIANDOLOetal97,SOLLICHetal97,BERTHIER-BARRAT,OHERNetal04,ILG-BARRAT07}. Much of our own most recent work is based on papers by Liu and colleagues, especially \cite{ONOetal02,HAXTON-LIU07}. The preceding authors showed that, in systems driven by external forces, slowly relaxing degrees of freedom may fall out of equilibrium with the thermal reservoir and, as a result, the temperature associated with those slow degrees of freedom differs from the reservoir temperature. More recently, the effective temperature has been used extensively in shear-transformation-zone (STZ) theories of large-scale plastic deformation in molecular glasses \cite{JSL04,BLP07I,BLP07II,JSL-MANNING-TEFF-07, EB-TEFF-PRE08, JSL-STZ-PRE08}. Perhaps its most remarkable success along these lines has been in explaining the nature of shear-banding instabilities in such systems \cite{SHI-FALK-SHEARBANDS-07,MANNINGetal-SHEARBANDS-08}.

Despite these successes, it has never been clear -- at least not to the present authors -- whether the effective temperature is truly a ``temperature'' in the thermodynamic sense, or whether it merely resembles a temperature in some aspects but differs in others.  We have argued, e.g. in \cite{JSL04,BLP07I,EB-TEFF-PRE08, JSL-STZ-PRE08}, that a noncrystalline material should be characterized in large part by its internal state of disorder, which we identify with the configurational entropy. Then the intensive quantity thermodynamically conjugate to this entropy, i.e. the derivative of the configurational energy with respect to the configurational entropy, must have the properties of a temperature in the sense that systems with higher temperatures are more highly disordered, and neighboring subsystems are in equilibrium with each other if they are both at the same temperature.

Our goal here is to make these ideas more precise. By doing so, we wish to determine what roles the effective temperature can play in dynamical theories of glassy phenomena, and what its limitations might be. Questions to be addressed include:  Under what circumstances is the effective temperature different from the ordinary reservoir temperature?  What are the forms of the first and second laws of thermodynamics in systems where both ordinary and effective temperatures are well defined?  What is the equation of motion for the effective temperature?  For the most part, our answers confirm the guesses made in earlier publications, but there are some differences.

To answer these questions, we have had to reexamine some basic principles of thermodynamics, especially the meaning of the second law in a non-standard situation of the kind we encounter here; and we also have had to understand the thermodynamic roles played by internal degrees of freedom in nonequilibrium situations. We have reported progress along those lines in the preceding paper \cite{EB-JSL-09-I}.  Much of the present analysis is based on the results described there.

In Sec. \ref{subsystems} of this paper, we introduce the hypothesis that a glassy system consists of two weakly coupled subsystems, one of which contains the slow configurational degrees of freedom that are characterized by the effective temperature. In order to focus as sharply as possible on the basic thermomechanical issues, as opposed to questions about spatial variations or time dependent orientations of the stress, we specialize to the case of a spatially uniform system.  We make several other simplifying assumptions, to be specified later, for the same reason. We derive the the two-temperature equations of motion in Sec. \ref{2T}, and conclude in Sec. \ref{conclusions} with remarks about the theory.

\section{Separable Subsystems}
\label{subsystems}

Our basic premise is that the degrees of freedom of a classical ($\hbar\!=\!0$) noncrystalline material, either below or not too far above a glass transition, can to a good approximation be separated into two parts: first, a set of configurational coordinates describing the mechanically stable positions of the molecules in their inherent states \cite{GOLDSTEIN69,STILLINGER-WEBER82,STILLINGER88}, and, second, the remaining kinetic and vibrational variables. We assume that the two sets of degrees of freedom describe weakly coupled subsystems of the system as a whole, in analogy to the usual thermodynamic analysis in which two neighboring subsystems of a larger system are brought into weak contact. In the latter situation, if the temperatures of the neighboring subsystems differ from each other, we know from the second law of thermodynamics that heat flows from the hotter to the colder subsystem; and we can compute the rate of heat flow either phenomenologically by defining a transport coefficient, or sometimes by computing that coefficient from first principles. We propose to repeat such an analysis for configurational and kinetic/vibrational degrees of freedom in weak contact with each other.

The configurational and kinetic/vibrational subsystems are distinguished  by their different time scales.  The kinetic/vibrational degrees of freedom are fast; they move at molecular speeds.  The configurational degrees of freedom are slow; they change on time scales no shorter than those associated with molecular rearrangements in glassy materials.  If these two time scales are comparable with one another, then the system is liquidlike, and the concept of an effective disorder temperature is irrelevant; but, as will be seen, the separation between fast and slow variables may remain useful.

To be more specific about the separability of these subsystems, consider a potential-energy landscape in $N d$-dimensional configuration space, where $N$ is the number of molecules and $d$ is the dimensionality.  The inherent states are the molecular configurations at the local energy minima in this space; and the configurational subsystem, apart from uniform translations and rotations, is defined by the denumerable set of these states.  A molecular rearrangement of the kind that occurs during plastic deformation corresponds to the motion of a system point from the near neighborhood of one inherent state to that of another.

If the total energy of the system lies below all but a small fraction of the energy minima, then the intersection of the potential-energy landscape with the surface of constant energy consists primarily of isolated, compact, manifolds enclosing the low-lying inherent states. Very few energy-conserving, deterministic trajectories connect the neighborhoods of different inherent states. For a large enough system that is not totally jammed, however, there always must exist some such trajectories; neighboring molecules always can move around each other if a large enough number of the other molecules move out of their way, as happens in the excitation-chain mechanism for example \cite{06Lan}. These connecting trajectories -- narrow filaments in configuration space -- occupy only a small fraction of the constant-energy manifold.  Thus, in physical space, they correspond to highly improbable, collective events.  The probabilities of such events are enhanced when the transitions are assisted by thermal fluctuations, which are not yet part of this picture but will enter shortly.

By definition, the motions of the kinetic/vibrational subsystem occur only on trajectories within the compact manifolds and not on the filamentary connections between them.  Our principal assumption is that the latter motions can accurately be approximated by weak, thermodynamically constrained, couplings between the configurational and kinetic/vibrational subsystems. That is, molecular rearrangements in the configurational subsystem -- transitions between inherent states -- are driven by noise generated in the kinetic/vibrational subsystem; conversely, heat generated during irreversible rearrangements flows through the kinetic/vibrational subsystem to a thermal reservoir.

We further assume that we can accurately approximate the total potential energy near any one of the minima by the inherent-state energy plus the harmonic energy of small-amplitude molecular vibrations about that state. Then, for a system of $N$ molecules in the neighborhood of the $\nu$'th inherent state, the total energy $H$ is given by
\begin{equation}
\label{Utotal}
H \cong H_C\{{\B r}_{\nu}\}+H_K\{{\B p}, \delta {\B r}_{\nu}\},
\end{equation}
where
\begin{equation}
\label{W}
H_K\{{\B p}, \delta {\B r}_{\nu}\} = K\{{\B p}\}+W\{\delta {\B r}_{\nu}\}.
\end{equation}
Here, $K\{{\B p}\}$ is the kinetic energy as a function of the set of $N$ momenta $\{{\B p}\}$, $H_C\{{\B r}_{\nu}\}$ is the inherent-state energy at the $\nu$'th minimum located at $\{{\B r}_{\nu}\}$, and $W\{\delta {\B r}_{\nu}\}$ is a quadratic form in the molecular displacements $\{\delta {\B r}_{\nu}\}$ measured from that minimum \cite{comment}. Note that the separation of the subsystems postulated in Eq. (\ref{Utotal}) is mathematically exact if the vibrational spectrum associated with $H_K\{{\B p}, \delta {\B r}_{\nu}\}$ were independent of the inherent-state index $\nu$; this is not quite true in general, but is a rather good approximation below the glass transition in the absence of irreversible volume deformation \cite{99SKT}. This issue will be discussed in detail in Sec. \ref{2T} below.

The configurational part of the internal energy, $H_C$, must depend on the size and shape of the system; i.e. $H_C$ must contain at least the macroscopic part of the elastic energy. Since, by construction, both subsystems  occupy the same region of physical space, we must exclude the macroscopic displacement modes from $H_K$.  Doing so is consistent with the separation of time scales because elastic modes that span the whole system have frequencies that are vanishingly small in the limit $N \to \infty$.  We thus assume that the bulk elasticity resides primarily in the configurational subsystem.  The kinetic/vibrational subsystem consists only of the high-frequency, localized modes that, in a disordered noncrystalline material, do not probe the boundaries of  the system and are sensitive only to the average spacing between the molecules.  Therefore, we assume that the kinetic/vibrational subsystem has an elastic bulk modulus and contributes a partial pressure to the system as a whole; but it has no shear modulus.

Note that we are maintaining the analogy between the present picture of weakly coupled configurational and kinetic/vibrational subsystems, and the usual thermodynamic picture of spatially neighboring subsystems. In the usual picture, the neighboring dynamical systems have coordinates in common only for the molecules that interact with each other across the surface of contact, and the coupling is weak because it involves only a small fraction of the total number of degrees of freedom.  Here, because we assume that the basins of attraction of the inherent states do not appreciably overlap or,  equivalently, that the vibrational amplitudes are small, the coupling is weak because only a small fraction of the configuration space contains trajectories that link different inherent-states.

\section{Two-Temperature Thermodynamics}
\label{2T}

\subsection{Internal Energy and Entropy}

In accord with the preceding discussion, we consider the deformation of a spatially uniform, amorphous system, in contact with a thermal reservoir at temperature $\theta_R$. We assume that $\theta_R$ is either below or not too far above the glass temperature $\theta_g$, so that the two-temperature decoupling can take place. As in \cite{EB-JSL-09-I}, we express temperatures in units of energy, and set Boltzmann's constant $k_B$ equal to unity.

Our goal of identifying temperature-like quantities requires that we start the statistical analysis by working in a microcanonical ensemble where the internal energy of each subsystem is a function of its entropy -- in principle, an independently computable quantity obtained by counting states. Thus we write the total, extensive, internal energy of the system, including a thermal reservoir, in the form
\begin{eqnarray}
\label{Ueqn}
U_{tot} \simeq U_C(S_C,{\B E}_{el},\{\Lambda_{\alpha}\}) + U_K(S_K,V_{el}) + U_R(S_R) \ ,\nonumber\\
\end{eqnarray}
where $U_C$ and $U_K$, respectively, are the configurational and kinetic/vibrational internal energies obtained from the energy functions $H_C$ and $H_K$ discussed in Sec. \ref{subsystems}, $S_C$ and $S_K$ are the respective entropies, and $\{\Lambda_{\alpha}\}$ denotes a set of internal state variables. Variations of the $\Lambda_{\alpha}$ describe irreversible changes in the state of the system which may be -- but are not necessarily -- coupled directly to shape deformations. $U_R$ is the energy of the thermal reservoir, which we assume to be strongly coupled to the kinetic/vibrational subsystem.

${\B E}_{el}$ is a symmetric elastic strain tensor that, in principle, includes both deviatoric (shear) and volumetric deformations. We maintain that only the reversible part of the strain can appear as an independent argument of the internal energy. The inelastic deformation can play no role here, because it must be defined as the displacement from some initial reference configuration and therefore is determined by the entire history of prior deformations. We insist that the motion of the system at any given time be determined entirely the external forces acting on it and the current values of the internal state variables. In other words, the memory of prior deformations is carried entirely by those state variables. The inelastic deformation itself can be determined only by integrating the equations of motion for those quantities.

In using the elastic part of the deformation as an independent argument of the internal energy, we are assuming that an elastic strain tensor can be identified unambiguously. This is a controversial assumption, for example, see the discussion in \cite{06XBM}; but it is not crucial for the main theme of this paper.  Therefore, for simplicity, we restrict ourselves to situations in which elastic displacements are small and linear elasticity is accurate; but we consider arbitrarily large inelastic deformations.

As argued in Sec. \ref{subsystems}, the kinetic/vibrational subsystem has a bulk modulus, and therefore the internal energy $U_K$ in Eq. (\ref{Ueqn}) can depend on the elastic part of the volume deformation, i.e. $V_{el}\!=\!V\!~tr{\B E}_{el}$; but it does not depend on ${\B E}_{el}$ itself. The absence of a shear modulus for this subsystem follows rigorously from our assumption of linear elasticity.  To see this,  invert $U_K(S_K,V_{el})$ to obtain the kinetic/vibrational entropy $S_K$ in the form
\begin{equation}
\label{S_K}
S_K = S_K(U_K,V_{el}) \ .
\end{equation}
In the harmonic approximation for $H_K$ in Eq. (\ref{Utotal}), we have
\begin{equation}
\label{S_K1}
S_K = 3\,N\,\left[1+\ln\left({U_K\over 3\,N}\right)\right]-\sum_{n=1}^{3N}\ln\,(\omega_n) \ ,
\end{equation}
where the $\omega_n$ are the frequencies of the vibrational modes, and we have assumed there to be $3 N$ kinetic and $3 N$ vibrational degrees of freedom. The dependence of $S_K$ on the deformation resides in the frequencies $\omega_n$, which are the eigenvalues of the dynamical matrix -- the second derivatives of $W\{\delta {\B r}_{\nu}\}$ in Eq. (\ref{W}). For linear elasticity, the $\omega_n$ are linear in that strain, and therefore the first-order shifts in the normal-mode frequencies must change sign when the shear strain changes sign. On the other hand, rotational and inversion symmetries in a statistically isotropic system preclude such sign changes.  Thus, the $\omega_n$ must be independent of the shear strain to first order, and the shear modulus must vanish.

The preceding discussion implies that the normal-mode frequencies $\omega_n$ depend on the total volume $V$, and not just on its elastic part. Thus, in principle, $U_K$ depends on the internal state variables.  This mechanical coupling between the kinetic/vibrational and configurational subsystems could be accounted for within the present formulation, but only at the expense of unnecessary complication. We therefore assume that the dependence on the set of internal state variables $\{\Lambda_{\alpha}\}$ resides exclusively in the configurational internal energy $U_C$.

In analogy to the expression for the internal energy in Eq. (\ref{Ueqn}), the total entropy is given by
\begin{eqnarray}
\label{Seqn}
S_{tot} \simeq S_C(U_C,{\B E}_{el},\{\Lambda_{\alpha}\})+ S_K(U_K,V_{el}) + S_R(U_R) \ .\nonumber\\
\end{eqnarray}
Note that the ``nonequilibrium'' subscript used in \cite{EB-JSL-09-I} is no longer needed here.  We assume that we can invert the expression for any one of these three entropies to obtain the corresponding internal energy function in Eq. (\ref{Ueqn}) or {\it vice versa}.

In principle, we have three separate temperatures: the effective temperature $\chi$, the kinetic/vibrational temperature $\theta$, and the reservoir temperature $\theta_R \cong \theta$. These are given respectively by
\begin{eqnarray}
\label{theta-U}
\chi = \left({\partial U_C \over \partial S_C}\right)_{{\B E}_{el},\{\Lambda_\alpha\}},~~\theta = \left({\partial U_K \over \partial S_K}\right)_{V_{el}},~~ \theta_R = {\partial U_R \over \partial S_R} \ .
\nonumber\\
\end{eqnarray}
Note that we depart here from the notation used in \cite{JSL-STZ-PRE08} and other earlier STZ papers, where $\chi$ was defined to be a dimensionless ratio of the effective temperature to an STZ formation energy in units of Boltzmann's constant.  Here, $\chi$ has the units of energy. The thermodynamic stresses are
\begin{equation}
\label{s-U}
V\,{\B T}_C = \left({\partial U_C\over \partial {\B E}_{el}}\right)_{S_C,\{\Lambda_\alpha\}}, \quad p_K = -\left({\partial U_K\over \partial V_{el}}\right)_{S_K} \ .
\end{equation}

\subsection{The First Law of Thermodynamics}

The first law of thermodynamics for this system is
\begin{equation}
\label{firstlaw1}
V\,{\B T}:{\B D} = \dot U_{tot},
\end{equation}
where the total stress $\B T$ is the sum of the partial stresses, ${\B T}={\B T}_C + {\B T}_K$. ${\B D}$ is the total rate of deformation tensor, including both elastic and inelastic parts. Note that all the stress quantities can be expressed as a sum $\B T=-p\,\bm{1} + {\B s}$ of hydrostatic ($p$) and deviatoric ($\B s$) contributions. Breaking Eq. (\ref{firstlaw1}) into its component parts, we obtain
\begin{equation}
\label{firstlaw1a}
(\dot U_C -V\,{\B T}_C:{\B D})+(\dot U_K - V\,{\B T}_K:{\B D})+\dot U_R = 0.
\end{equation}

To evaluate these terms, look first at the configurational subsystem, and recall that changing the shape of the system at fixed values of $S_C$ and $\{\Lambda_{\alpha}\}$ is by definition an ``elastic'' deformation; there is no irreversible change in the internal state of the system. Using this definition, we have
\begin{equation}
\label{dUCdt-el}
\left({\partial U_C\over \partial t}\right)_{S_C,\{\Lambda_\alpha\}} = V\,{\B T}_C:{\B D}_{el},
\end{equation}
where ${\B D}_{el}$ is the elastic part of the rate of deformation tensor. Because we assume that the total rate of deformation is a sum of elastic and inelastic parts,
\begin{equation}
{\B D}= {\B D}_{el}+ {\B D}_{in},
\end{equation}
the configurational component of Eq. (\ref{firstlaw1a}) is
\begin{eqnarray}
\nonumber
\dot U_C &-&V\,{\B T}_C:{\B D} = \chi\,\dot S_C -V\,{\B T}_C:{\B D}_{in}\cr\\ &+& \sum_{\alpha} \left({\partial U_C\over \partial \Lambda_{\alpha}}\right)_{S_C,{\B E}_{el}}\dot \Lambda_{\alpha}.
\end{eqnarray}

Next, consider the kinetic/vibrational component of Eq. (\ref{firstlaw1a}), where
\begin{equation}
\label{dUKdt}
\left({\partial U_K\over \partial t}\right)_{S_K} = -p_K \dot{V}_{el} \ .
\end{equation}
Therefore,
\begin{equation}
\label{1stKsub}
\dot U_K -\,V\,{\B T}_K:{\B D} = \theta\,\dot S_K - \,V\,{\B s}_K:{\B D}^{dev} + p_K \dot{V}_{in},
\end{equation}
where $\B D^{dev}$ is the deviatoric (non-hydrostatic) part of the total rate of deformation $\B D$, and we have used $\dot V = \dot V_{el} + \dot V_{in}$ to eliminate $\dot V_{el}$. Note the appearance of $\B s_K$ in the last equation. If it has no thermodynamic deviatoric stress, cf. Eq. (\ref{s-U}), then the kinetic/vibrational subsystem can support a viscous shear stress, which can appear in Eq. (\ref{1stKsub}) in the presence of nonzero shear flow, i.e. when ${\B D}^{dev}\!\ne\!0$.

Recombining terms, we find that the first law, Eq. (\ref{firstlaw1}), becomes
\begin{eqnarray}
\label{firstlaw2}
\chi\,\dot S_C &+& \theta\,\dot S_K + \dot U_R - V\,{\B T}_C:{\B D}_{in}+ p_K \dot{V}_{in}\\
\cr &-& V\,{\B s}_K:{\B D}^{dev} +\sum_{\alpha} \left({\partial U_C\over \partial \Lambda_{\alpha}}\right)_{S_C,{\B E}_{el}}\dot \Lambda_{\alpha}= 0 \ .\nonumber
\end{eqnarray}
Using
\begin{equation}
- V\,{\B T}_C:{\B D}_{in} = - V\,{\B s}_C:{\B D}^{dev}_{in}+ p_C \dot{V}_{in},
\end{equation}
we rewrite Eq. (\ref{firstlaw2}) in the form
\begin{eqnarray}
\label{firstlaw3}
\chi\,\dot S_C &+& \theta\,\dot S_K + \dot U_R - V\,{\B s}_C:{\B D}^{dev}_{in}+ p \dot{V}_{in}\\
\cr &-& V\,{\B s}_K:{\B D}^{dev} +\sum_{\alpha} \left({\partial U_C\over \partial \Lambda_{\alpha}}\right)_{S_C,{\B E}_{el}}\dot \Lambda_{\alpha}= 0 \ .\nonumber
\end{eqnarray}
Note that the partial deviatoric stress $\B s_C$ couples to the deviatoric, inelastic rate of deformation ${\B D}^{dev}_{in}$, the total pressure $p$ couples to the inelastic part of the volume rate of change $\dot{V}_{in}$, and the partial deviatoric stress $\B s_K$ couples to the total deviatoric rate of deformation ${\B D}^{dev}$. These features will reappear below in dissipation inequalities based on the second law of thermodynamics.

\subsection{The Second Law of Thermodynamics}

The second law of thermodynamics requires that the total entropy be a non-decreasing function of time
\begin{equation}
\label{secondlaw}
\dot S_C + \dot S_K + \dot S_R \ge 0,
\end{equation}
where
\begin{equation}
\dot U_R = \theta_R\,\dot S_R.
\end{equation}
Use Eq. (\ref{firstlaw3}) to eliminate $\dot S_C$ in Eq. (\ref{secondlaw}), and rearrange the terms as follows
\begin{eqnarray}
\label{secondlaw2}
\nonumber
&-&\!\!\left(1 - {\theta\over \theta_R}\right)\,\dot U_R - \left(1-{\chi\over \theta}\right)\left(\theta\,\dot S_K + {\theta\over \theta_R}\,\dot U_R\right) \\
\cr &+&\!\! V\,{\B s}_K:{\B D}^{dev} + {\cal W}_C({\B s}_C,p,\{\dot\Lambda_\alpha\})\ge 0,
\end{eqnarray}
where
\begin{eqnarray}
\label{calWdef}
&&{\cal W}_C({\B s}_C,p,\{\dot\Lambda_\alpha\}) = V\,{\B s}_C:{\B D}^{dev}_{in}-p\dot{V}_{in}\nonumber\\
\cr &&-\sum_{\alpha} \left({\partial U_C\over \partial \Lambda_{\alpha}}\right)_{S_C,{\B E}_{el}}\dot \Lambda_{\alpha}
\end{eqnarray}
where ${\cal W}$ is the analog of the quantity defined in Eq. (4.11) of \cite{EB-JSL-09-I}.
Note that ${\cal W}_C$ is the difference between the rate at which inelastic work is being done on the configurational subsystem and the rate at which energy is stored in the internal degrees of freedom. As in \cite{EB-JSL-09-I}, it is an energy dissipation rate.

The left-hand side of the inequality in Eq. (\ref{secondlaw2}) is the sum of independently variable quantities that must separately be non-negative if the inequality is to be satisfied for all possible motions of the system. Thus, in the spirit of Coleman and Noll \cite{COLEMAN-NOLL-63}, we require that
\begin{equation}
-\left(1 - {\theta\over \theta_R}\right)\,\dot U_R \ge 0 \ ,
\end{equation}
\begin{equation}
- \left(1-{\chi\over \theta}\right)\left(\theta\,\dot S_K + {\theta\over \theta_R}\,\dot U_R\right)\ge 0 \ ,
\end{equation}
\begin{equation}
{\B s}_K:{\B D}^{dev}\ge 0 \ ,
\end{equation}
and
\begin{equation}
\label{WCineq}
{\cal W}_C({\B s}_C,p,\{\dot\Lambda_\alpha\})\ge 0 \ .
\end{equation}

The first two of these inequalities pertain to the heat flow between the subsystems. We have arranged them in a way that reflects the fact that the thermal coupling between the configurational and kinetic/vibrational subsystems is expected to be weak, and the coupling between the kinetic/vibrational subsystem and the reservoir should be strong. To satisfy them, we write
\begin{equation}
\label{Adef}
\theta\,\dot S_K + {\theta\over \theta_R}\,\dot U_R = -\,A(\chi,\theta)\,\left(1-{\chi\over \theta}\right) \ ,
\end{equation}
\begin{equation}
\label{heat_res}
\dot U_R = -\, B(\theta,\theta_R)\frac{\theta_R}{\theta}\,\left(1 - {\theta\over \theta_R}\right) \ ,
\end{equation}
where $A$ and $B$ are non-negative thermal transport coefficients. Using the definition of the kinetic/vibrational (extensive) heat capacity at constant volume
\begin{equation}
\theta\,\dot S_K = C_V^{K}\,\dot \theta
\end{equation}
and eliminating $\dot U_R$ between Eqs. (\ref{Adef}) and (\ref{heat_res}), we obtain the following heat equation for the kinetic/vibrational subsystem
\begin{equation}
\label{heat_k/v}
C_V^{K}\,\dot \theta = \, B(\theta,\theta_R)\,\left(1 - {\theta\over \theta_R}\right) -\,A(\chi,\theta)\,\left(1-{\chi\over \theta}\right) \ .
\end{equation}
The first term on the left hand side describes heat transfer between the kinetic/vibrational subsystem and the thermal reservoir, and the second term describes heat transfer between the kinetic/vibrational subsystem and configurational subsystem. Our assumptions imply that $A(\chi,\theta)$ is small because it contains the physics of the weak coupling described in Sec. \ref{subsystems}. On the other hand, a large value of $B$ implies that $\theta \to \theta_R$ such that
\begin{equation}
\label{Q_R}
Q_R \equiv \lim_{\substack{B\rightarrow \infty\\\theta\rightarrow \theta_R}} \, B(\theta,\theta_R)\,\left(1 - {\theta\over \theta_R}\right)
\end{equation}
is finite.  In this way, the thermal reservoir controls the temperature of the kinetic/vibrational subsystem.

The third term on the left hand side of Eq. (\ref{secondlaw2}) must, by itself, be non-negative for arbitrary values of the deviatoric part of the rate of deformation tensor, ${\B D}^{dev}$. We satisfy this requirement by assuming a linear relation of the form
\begin{equation}
\label{etadef}
{\B s}_K = \eta\,\B D^{dev},
\end{equation}
where the shear viscosity $\eta$ is a scalar for our isotropic material.

The fourth inequality, Eq. (\ref{WCineq}), pertains to the irreversible dynamics of the configurational subsystem. This inequality, which is a generalization of the standard Clausius-Duhem inequality \cite{MAUGIN-99} to the two-subsystems situation, must be satisfied by any constitutive model. The implications of this inequality for the STZ theory of amorphous plasticity \cite{JSL04,BLP07I,BLP07II,JSL-MANNING-TEFF-07, EB-TEFF-PRE08, JSL-STZ-PRE08} are discussed in the following paper \cite{EB-JSL-09-III}.

\subsection{Equation of Motion for $\chi$}

The crux of our two-temperature analysis is an equation of motion for the effective temperature $\chi$.  This is a heat-balance equation, which we obtain by rewriting the first law of thermodynamics as stated in Eq.(\ref{firstlaw3}). Define the effective (extensive) heat capacity at constant volume
\begin{equation}
\chi\,\dot S_C = C_V^{e\!f\!f}\,\dot \chi \ .
\end{equation}
Then use this definition and Eq. (\ref{Adef}) in Eq. (\ref{firstlaw3}) to obtain
\begin{eqnarray}
\label{firstlaw4}
\nonumber
&&C_V^{e\!f\!f}\,\dot \chi =
{\cal W}_C({\B s}_C,p,\{\dot\Lambda_\alpha\})\cr \\ &&+ V\,\eta\,\B D^{dev}:\!{\B D}^{dev} + A(\chi,\theta)\,\left(1-{\chi\over \theta}\right).
\end{eqnarray}
Here
\begin{equation}
\label{Qdef}
-A(\chi,\theta)\,\left(1-{\chi\over \theta}\right)\approx \theta\,\dot S_K + \dot U_R \equiv  - Q
\end{equation}
is the rate at which heat is flowing to the kinetic/vibrational subsystem and the thermal reservoir. The approximation in Eq. (\ref{Qdef}) is accurate when $\theta \!\approx\! \theta_R$ and the kinetic/vibrational subsystem serves as the thermal reservoir. Moreover, if the thermal reservoir has a large heat capacity (as is commonly assumed), Eq. (\ref{heat_k/v}) implies
\begin{equation}
\dot{\theta}=0,\quad Q_R= Q \ .
\end{equation}
Under these circumstances, Eq. (\ref{firstlaw4}) is a complete description of the energy flow in this system.

In the absence of deformation, i.e. when the term proportional to $A(\chi,\theta)$ is all that remains on the right-hand side of Eq. (\ref{firstlaw4}), or when $A(\chi,\theta)$ is large, then $\chi \to \theta$ and our two-temperature formulation reverts to an ordinary  one-temperature theory.  A single heat equation for $\chi = \theta$ can be obtained by summing Eqs. (\ref{heat_k/v}) and (\ref{firstlaw4}), and identifying $C_V\!=\!C^K_V\!+\!C_V^{e\!f\!f}$ as the total heat capacity.  For example, using this procedure, we recover the one-temperature equations of motion for viscoelastic volume deformations derived in \cite{EB-JSL-09-I}.

We reiterate the physical meaning of the different terms in Eq. (\ref{firstlaw4}). $C_V^{e\!f\!f}\,\dot \chi$ is the rate of change of the heat of disorder.  The first term on the right-hand side, ${\cal W}_C({\B s}_C,p,\{\dot\Lambda_\alpha\})$, is the non-negative rate of energy dissipation in the configurational subsystem, which we find to be the difference between the rate at which inelastic work is being done, ($V{\B s}_C\!:\!{\B D}^{dev}_{in}\!-\!p\dot{V}_{in}$), and the rate at which energy is being stored in the configurational degrees of freedom $\{\Lambda_{\alpha}\}$.   The term $V\,\eta\,\B D^{dev}\!:\!\B D^{dev}$ is the rate of viscous heating in the kinetic/vibrational subsystem.  Finally, $-A(\chi,\theta)(1-\chi/\theta)$ is the rate at which heat flows out of the configurational subsystem.

\section{Concluding Remarks}
\label{conclusions}

The main goal of this paper has been to derive the constraints imposed by the first and second laws of thermodynamics on driven, glass-forming systems whose states of configurational disorder can be characterized by an effective temperature, and whose nonequilibrium dynamics can be described by internal state variables. We stress that we have provided only a general thermodynamic description, and not yet a detailed physical theory of such systems. Although we have made several inessential simplifying assumptions -- spatial homogeneity and a simple distinction between elastic and inelastic deformations, for example -- our only fundamental physical assumption is weak coupling between the configurational and kinetic/vibrational subsystems.

To make further progress, we must make more specific physical assumptions. In particular, we must specify physical ingredients of the dissipation function ${\cal W}_C({\B s}_C,p,\{\dot\Lambda_\alpha\})$ defined in Eq. (\ref{calWdef}), and the thermal coupling coefficient $A(\chi,\theta)$ defined in Eq. (\ref{Adef}); and we must make use of the second-law constraints in both cases. As was shown for a special situation in \cite{EB-JSL-09-I}, non-negativity of  ${\cal W}_C({\B s}_C,p,\{\dot\Lambda_\alpha\})$ is a generalization of the Clausius-Duhem inequality. Inequalities of this kind are usually satisfied \cite{MAUGIN-99} by assuming the existence of variational principles such that the internal variables $\{\Lambda_\alpha\}$ move downhill in some free energy landscape. We shall see in \cite{EB-JSL-09-III} that, when we interpret the internal state variables $\{\Lambda_\alpha\}$ to be the STZ dynamical variables, we no longer recover a simple variational form of this inequality, and that the dynamical structure of the theory is much richer than it was in \cite{EB-JSL-09-I}. Similarly, much of the physics of the STZ theory resides in the coupling coefficient $A(\chi,\theta)$. It is here where the assumption of weak coupling between the subsystems must be described in more detail, albeit still in a phenomenological manner.

\begin{acknowledgments}
JSL acknowledges support from U.S. Department of Energy Grant No. DE-FG03-99ER45762.
\end{acknowledgments}


\begin{thebibliography}{99}
\bibitem{comment} Anharmonic contributions can be included, but they play no important role in the present context.
\end{thebibliography}
\end{document}